\documentclass[12pt]{article}
\usepackage{bm}
\usepackage{amsmath}
\usepackage{amssymb}
\usepackage{graphicx}
\usepackage{amsmath}
\usepackage{bbm}
\usepackage{comment}

\begin{document}
\makeatletter
\@addtoreset{equation}{section}
\makeatother
\renewcommand{\theequation}{\thesection.\arabic{equation}}
\baselineskip 15pt

\title{\bf Comments on a recent proposal of superluminal communication}

\author{GianCarlo Ghirardi\footnote{e-mail: ghirardi@ts.infn.it}\\ {\small
Emeritus, Department of  Physics, the University of Trieste,}\\ {\small the Abdus
Salam International Centre for Theoretical Physics, Trieste.}  }

\date{}

\maketitle

\begin{abstract}

\noindent We analyze critically a recent proposal of faster than light communication.
\end{abstract}



\section{Critical discussion of the proposal}
Recently R. Cornwall, with reference to a paper \cite {RC1} he wrote in 2006, has argued \cite{RC2} that the No-signalling theorem \cite{{ggrw},{grw},{h}} derived by the present author and his collegues\footnote{Actually we have written various other papers \cite {{gw},{gb},{gr}}  on the same argument, covering more general cases such as those involving non ideal measurements, which the author does not mention.} and by M. Hall,  is not correct \footnote{This  assertion is rather peculiar. In fact, Cornwall literarily writes: {\it An interferometer set up(1), utilizing entangled particles to receive information, by remote change in quantum state, by a modulator, has been criticized as not obeying the laws of quantum mechanics in that local changes in quantum state cannot affect remote physics(2-4)}. We note that the paper, ref.$(1)$, of Cornwall has been written in 2006, while the three papers - Cornwall refs.$(2-4)$ - which are claimed to criticize it have been written in the years 1987, 1988 and 1980, respectively. It is one of the  mysteries of \cite{RC2} how somebody might criticize a paper more or less 18-26 years before the paper itself was written.}
, since  ``it contains an omission or restriction in logic''.

In this paper we want to stress three important facts:
\begin{itemize}
\item That the arguments of refs.\cite {{RC1},{RC2}}, and, as a consequence those of  more recent papers \cite{{RC3},{RC4}} by the same author, are  basically flawed,
\item That the arguments developed by us in refs \cite{{ggrw},{grw}}, as well as in many other papers \cite {{gw},{gb},{gr}} we wrote on this subject, are  correct,
\item That Cornwall does  not  deal consistently with a quantum problem as simple as the  one he is considering.  
\end{itemize}

The first step which must be done is to summarize the content of refs.\cite{{RC1},{RC2}}. In them, the author considers an experiment which should lead to faster than light communication between far away regions. The proposal  makes resort, as almost all which appeared in the literature, to the instantaneous reduction at-a-distance of the wave packet. It is summarized synthetically in Fig.2 of Ref.[1] which we reproduce here for clarity sake.  A source S emits a pair of entangled photons propagating in opposite directions along the z-axis in the typical state  used in EPR-like set-ups with photons:

\begin{equation}
\vert\Phi\rangle=\frac{1}{\sqrt{2}}[\vert H\rangle_{1}\otimes\vert V\rangle_{2}+\vert V\rangle_{1}\otimes\vert H\rangle_{2}].
\end{equation}

Very far from the source in the direction (at left) of propagation of photon 1, there is a polarizing filter which is used to modulate the measurement performed in this region. The experimenter controlling the filter can choose to do nothing to the photon arriving there - a case denoted as (bit 0) - in which case the state of the composite system remains the one of Eq.(1.1), or, alternatively - (bit 1) - he can perform a ``modulation'' by measuring the vertical or horizontal polarization of photon 1, in which case, due to wave packet reduction, the state of the composite system becomes either $|H_{1}\rangle\otimes|V_{2}\rangle$ or $|V_{1}\rangle\otimes|H_{2}\rangle$ with equal probability.

 \begin{figure}[h]
\begin{center}
\includegraphics{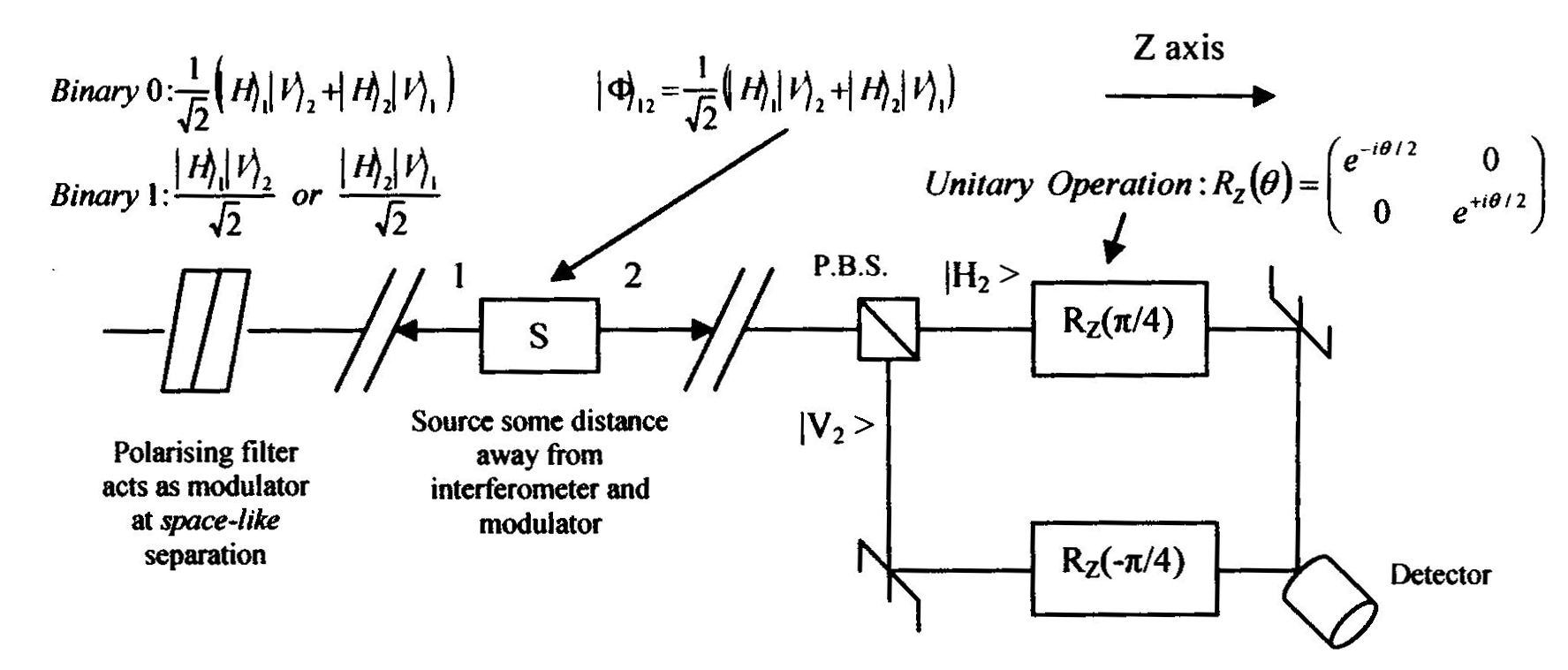}
\caption{Cornwall's proposal} 
\end{center}
\end{figure}

 Now let us pay attention to the Appendix 1 of Ref.[2] (which the author   has recently presented once more on research gate)  which has  the suggestive title ``Against the No-signalling Theorem''. It is just in this appendix that it is claimed  that our proofs \cite{{ggrw},{grw}} of the impossibility of sending faster than light signals by the reduction of the wave packet  contain an omission or restriction in logic. This statement is not proved at all in the mentioned papers by Cornwall, and the following calculations by him, aimed to show that superluminal signaling can actually be achieved,  are very confused and  basically wrong.
 
 In the just mentioned Appendix  the author considers, first of all, the case in which the distant observer, let me denote him as 1, performs a measurement aimed to check whether photon 1 is polarized horizontally or vertically, i.e., he takes into account the situation corresponding to (bit 1). 
 
 As already mentioned, if one respects the quantum rules, in the considered case reduction takes place either to the  state $|H_{1}\rangle\otimes|V_{2}\rangle$ or to the state $|V_{1}\rangle\otimes|H_{2}\rangle$ for the composite system. Since the two events are equally probable, to go on with the analysis, one must assume that the state has been transformed into a statistical mixture of the two just mentioned states. Correspondingly the statistical operator of the composite system in the Hilbert space which is the direct product of those of the two constituents has the form:
 
 \begin{equation}
\rho(bit 1;1,2)= \frac{1}{2}\{[\vert H_{1}\rangle\otimes\vert V_{2}\rangle\cdot\langle V_{2}\vert\otimes\langle H_{1}\vert]+[\vert V_{1}\rangle\otimes\vert H_{2}\rangle\cdot\langle H_{2}\vert\otimes\langle V_{1}\vert]\}.
\end{equation}

 Since the author is interested in the results of experiments made on photon 2, i.e., fundamentally, to what happens in the region where system 2 is located, he correctly claims that to deal with such experiments one must take the partial trace on system 1 getting in this way the statistical operator describing the physics of  system 2. This operation leads to the following statistical operator for such a system:
 
 \begin{equation}
\rho(bit 1;2)=\frac{1}{2}\{\vert V_{2}\rangle\cdot\langle V_{2}\vert+\vert H_{2}\rangle\cdot\langle H_{2}\vert\}.
\end{equation}

Concerning this part of the paper, we agree perfectly with the conclusions of the author, even though we cannot avoid to stress that, in the quoted Appendix, he makes a certain mess; for instance, he writes the ``states'' of the composite system as 2x2 matrices whose elements are state vectors, an absolutely  meaningless mathematical notation. However, as already stated, the conclusion is correct.
 
 We now pass to analyze the (bit 0) case, i.e. the one in which, at left, no measurement is performed before the photon at right impinges on the device which is located there.  Let us analyze the process and the corresponding mathematical description step by step. The state of the composite system, in the considered case and before photon 2 impinges on the  beam splitter which appears at right, is the initial state:
 
 \begin{equation}
\vert\Phi\rangle=\frac{1}{\sqrt{2}}[\vert H\rangle_{1}\otimes\vert V\rangle_{2}+\vert V\rangle_{1}\otimes\vert H\rangle_{2}].
\end{equation}

Photon 2, when it is in the state $|V_{2}\rangle$, by crossing the first beam splitter is, first of all,  deflected in the down direction, then deflected again and sent through  the phase shifter $R_{z}(-\pi/4)$, which attaches an appropriate phase to the state \footnote{Here, obviously, we pay attention to the spin states, since the specification of the other variables of the photon can be safely ignored.}, i.e., the impinging state $|V_{2}\rangle$ becomes  $[e^{i\theta}|V_{2} \rangle]$.

For what concerns the second term, the one referring to photon 2 of the initial state (1.4), since its polarization is horizontal it is transmitted undeflected by the beam splitter, then it traverses the phase shifter $R_{z}(\pi/4)$, which attaches an appropriate phase to it \footnote{ Here we have resorted to a very sketchy argument concerning the phases. Actually the phases attached to the states might have different values from the indicated ones due to the specific   working of the beam splitter itself or to the inclusion of appropriate delay terms. However, it has to be stressed that all our  subsequent arguments have nothing to do with the specific values of the phases, which might take arbitrary different values at the experimenter's whim.} so that  the impinging state $|H_{2}\rangle$ becomes  $[e^{-i\theta}|H_{2} \rangle]$. 

Accordingly, the state for the composite system we have to deal with is:

 \begin{equation}
\vert\Phi^{*}\rangle=\frac{1}{\sqrt{2}}[e^{i\theta}\vert H\rangle_{1}\otimes\vert V\rangle_{2}+e^{-i\theta}\vert V\rangle_{1}\otimes\vert H\rangle_{2}].
\end{equation}

Since we are interested in the physics of photon 2, and the considered state refers to the composite system of two photons, we must now to go through the standard procedure of building the statistical operator for the composite system, $\rho(bit 0;1,2)=|\Phi^{*}\rangle\langle\Phi^{*}|$ and to take the partial trace of it on the degrees of freedom of photon 1.

We have, first of all:
\begin{eqnarray}
\rho(bit 0;1,2)&=&|\Phi^{*}\rangle\langle\Phi^{*}|=\frac{1}{2}[|H_{1}>|V_{2}><V_{2}|<H_{1}|+|V_{1}>|H_{2}><H_{2}|<V_{1}| \nonumber \\
&+&e^{2i\theta}|H_{1}>|V_{2}><H_{2}|<V_{1}|+e^{-2i\theta}|V_{1}>|H_{2}><V_{2}|<H_{1}|].
\end{eqnarray}

We have chosen to make explicit all four terms characterizing the statistical operator of the composite system to put in clear evidence that, while the first two terms contain kets and bras referring to the same state of photon 1, the other two terms contain a ket different from (and orthogonal to the one of) the bra. Now we must take the partial trace on system 1 to have the statistical operator describing system 2. The trace is obtained by summing the diagonal elements in an {\bf arbitrary} ortonormal basis, so that it is particularly convenient to use as such a basis precisely the  complete set (in the spin space of photon 1) of the ortogonal states $|H_{1}>$ and $|V_{1}>$. The result is obvious: the statistical operator describing system 2 in the case in which at left one chooses (bit 0) turns out to be simply:

\begin{equation}
\rho(bit 0; 2)=\frac{1}{2}\{\vert V_{2}\rangle\cdot\langle V_{2}\vert+\vert H_{2}\rangle\cdot\langle H_{2}\vert\},
\end{equation}
\noindent which is precisely the same (Eq. (1.3)) of the case in which (bit 1) has been chosen at left.

Apart from the trivial nature of the argument we would like to stress that the author uses the quantum formalism in a wrong way, and to call the attention of the reader on the fact that he has not grasped the absolutely elementary but important fact that while a superposition of states of a single particle, like $\frac{1}{\sqrt{2}}[|V_{2}>+|H_{2}>]$, when sent through a Mach-Zender-like interferometer leads to interference effects which can be tuned, such effects cannot appear (for system 2) when the two states appearing in the previous expression are associated to different states of another system, so that the state of the composite system is entangled.

All what I have said can be formalized in the following almost trivial theorem which coincides with the one we have presented \footnote{ A theorem  about which A. Shimony \cite{s} has written that it represents  ``the first rigorous proof in the literature that faster than light signalling is not possible within quantum mechanics''.} in ref.[4]:

\begin{itemize}
\item The effect of a measurement on one (let us assume it is the one labeled as 1) of the constituents of a composite (1+2) system must be described, in the composite system language, by:
\begin{equation}
\rho(1,2)\rightarrow \rho^{(*)}(1,2)= \sum_{i} P^{(1)}_{i}\rho(1,2)P^{(1)}_{i},
\end{equation}
\noindent where the $P^{(1)}_{i}$ are the orthogonal projection operators associated to the eigenmanifolds of the measured observable,
\item The trace operation (over the Hilbert space of system 1) enjoys of the cyclic property concerning observables of the same space, and the projection operators $P^{(1)}_{i}$ are idempotent, so that:
\begin{eqnarray}
&&Trace^{(1)}\rho^{(*)}(1,2)=Trace^{(1)}[ \sum_{i} P^{(1)}_{i}\rho(1,2)P^{(1)}_{i}]=\\ \nonumber &=&Trace^{(1)}[ \sum_{i} P^{(1)2}_{i}\rho(1,2)]=Trace^{(1)}[ \sum_{i} P^{(1)}_{i}\rho(1,2)]=Trace^{(1)}[ \rho(1,2)],
\end{eqnarray}
\noindent where we have taken into account the completeness, $\sum_{i}P^{(1)}_{i}=I^{(1)}$, of the set $\{P^{(1)}_{i}\}$ of the projection operators. Accordingly, to perform or not the measurement process on photon 1 is totally irrelevant for the physics of photon 2.

\end{itemize}

It goes without saying that since we have proved once more that no faster than light signal can be sent by doing any legitimate (see our other papers on the subject which cover all possibilities) action  on one of a pair of entangled systems, the whole argument of Refs.[1,2] breaks down and, accordingly,  there is no reason to analyze it in greater detail and to discuss its compatibility  with relativistic requirement and the possibility of synchronizing far away clocks as the author does in the considered and in the other  papers he has written on this subject.

\end{document}